\documentclass[aps,preprint,groupedaddress]{revtex4-1}
\usepackage{graphicx}
\usepackage{dcolumn}
\usepackage{bm}

\begin{document}

\title{Hot electron transport in a strongly correlated transition metal oxide}

\author{Kumari Gaurav Rana$^1$, Takeaki Yajima$^{2,3}$, Subir Parui$^1$, Alexander F. Kemper$^3$, Thomas P. Devereaux$^3$, Yasuyuki Hikita$^3$, Harold Y. Hwang$^{3,4}$}
\author{Tamalika Banerjee$^{1,}$}
 \email{T.Banerjee@rug.nl}

\affiliation{$^1$
Physics of Nanodevices, Zernike Institute for Advanced Materials, University of Groningen, The Netherlands
}
 \affiliation{$^2$
Department of Materials Engineering, The University of Tokyo, Bunkyo-ku, Tokyo, 113-8656, Japan
}
 \affiliation{$^3$
Stanford Institute for Materials and Energy Sciences, SLAC National Accelerator Laboratory, Menlo Park, California 94025, USA
}
\affiliation{$^4$
Geballe Laboratory for Advanced Materials, Department of Applied Physics, Stanford University, Stanford, California 94305, USA
}

\date{\today}

\maketitle

\footnotesize \indent {\bf Oxide heterointerfaces are ideal for investigating strong correlation effects to electron transport, relevant for oxide-electronics. Using hot-electrons, we probe electron transport perpendicular to the La$_{0.7}$Sr$_{0.3}$MnO$_{3}$ (LSMO)- Nb-doped SrTiO$_3$ (Nb:STO) interface and find the characteristic hot-electron attenuation length in LSMO to be 1.48  $\pm$ 0.10 unit cells (u.c.) at -1.9 V, increasing to 2.02 $\pm$ 0.16 u.c. at -1.3 V at room temperature. Theoretical analysis of this energy dispersion reveals the dominance of electron-electron and polaron scattering. Direct visualization of the local electron transport shows different transmission at the terraces and at the step-edges.}\\ 

\indent Heterointerfaces between strongly correlated transition-metal oxides have proven to be ideal platforms for investigating new physical phenomena in condensed-matter and for designing multifunctional devices for oxide electronics \cite{HwangTokura}. Increasingly, many of the exciting device prospects with such materials involve hot-electron transport, such as in photovoltaic effects in multiferroics, manganite transistors, ferroelectric tunnel-junctions, etc.\cite{CheongPV, RameshNatNanotech2010, Takeaki, Garcia2009, Maksymovych}. Hot electrons, characterized by an energy higher than the Fermi energy, $E_F$, by more than a few times the thermal energy, are an interesting probe to investigate the physics of electron transport in different material systems. At such energies, the fundamental scattering processes are different than at $E_F$ and include elastic/quasielastic scattering, and inelastic scattering via electron, phonon and spin wave excitations. Experimental techniques and devices that exploit hot electron transport are found in a variety of electron spectroscopy techniques\cite{Aeschlimann}, in spintronic devices such as the spin-valve transistor\cite{MonsmaScience}, in spin-transfer torque devices\cite{BuhrmanSankey}, in Si spin injection devices\cite{AppelbaumNature}, and in graphene based optoelectronic devices\cite{Gabor}, and have yielded crucial insights into the transport properties in this energy regime.\\
\indent In this context, very little is known about hot-electron transport in oxide heterointerfaces with transition-metal oxides, particularly in the presence of strong correlations between the electron's charge, spin and orbital degrees of freedom\cite{HwangTokura}. Such interfaces are also attractive for designing  multifunctional devices that do not necessarily scale according to Moore's law. Here we address this by using the technique of Ballistic Electron Emission Microscopy (BEEM)\cite{beem} and probe hot-electron transport across an archetypal oxide ferromagnet La$_{0.7}$Sr$_{0.3}$MnO$_{3}$ (LSMO) on n-type semiconducting Nb-doped SrTiO$_3$ (Nb:STO). We find the characteristic hot-electron attenuation length in LSMO to be 1.48  $\pm$ 0.10 unit cells (u.c.; 1 u.c. = 0.39 nm) at -1.9 V, increasing to 2.02 $\pm$ 0.16 u.c. at -1.3 V at room temperature. Theoretical analysis of this dispersion reveals the dominance of electron-electron and polaron scattering at these energies. \\ 
\indent In an oxide heterojunction, electrical transport has been commonly studied using a Schottky diode involving unconventional semiconductors, often derived by doping Mott or band insulators. Probing electron transport using a Schottky interface with transition metal oxides has provided useful insights into the band bending, band offsets and their sensitivity to interface states, chemical doping and external magnetic and electric fields\cite{HikitaBook}. However, the contribution of long range correlation effects to the transport of electrons (depth-resolved) and quantification of transport parameters such as the hot electron attenuation length, carrier lifetime, etc., in transition metal oxides has not been explored. Such studies involving pervoskite metal-semiconductor (M-S) interfaces are important as they are the building blocks of most oxide electronic devices.\\
{\bf Results}\\
\indent We use a current-perpendicular-to-plane (CPP) configuration to probe vertical transport of hot electrons in the metallic oxide ferromagnet LSMO, across an epitaxial Schottky interface with Nb:STO, using the versatile technique of BEEM\cite{beem}. In addition to providing direct quantification of the energy dependent hot electron attenuation length, such a study using BEEM also enables us to analyse nanoscale spatial inhomogeneity of the transport in buried layers with high lateral resolution\cite{vonkanel1}. For these studies, we use 0.01 wt. $\%$ Nb:STO (001) substrates and deposit LSMO of variable thickness. The devices were fabricated by pulsed laser deposition using TiO$_2$-terminated Nb:SrTiO$_3$ (001) substrates (Nb = 0.01 wt. $\%$). A single unit cell of SrMnO$_3$  was first grown to enhance the Schottky Barrier Height and suppress reverse bias leakage \cite{HikitaPRBSBH}, and subsequently LSMO films were grown at the O$_2$ partial pressure of 10$^{-1}$ Torr, substrate temperature of 850$^{\circ}$ C, and at a laser fluence of 0.8 J/cm$^{2}$. The deposited layer thicknesses were controlled by using reflection high-energy electron diffraction intensity oscillations. For Ohmic contacts, gold was evaporated onto the LSMO, and indium was ultrasonically soldered onto the Nb:STO. In BEEM, a three-terminal transistor configuration, the top LSMO surface is grounded and  a negative voltage, \textit{V$_T$}, is applied to the Scanning Tunneling Microscope (STM) tip, with the tunnel current, \textit{I$_T$}, kept constant using feedback (Fig. 1).  A modified commercial Ultra High Vaccuum (UHV) STM system from RHK Technology was used for the BEEM studies and the measurements were performed at 300 K using PtIr metal tips. The maximum kinetic energy of the electrons injected from the STM tip is given by the applied bias \textit{e}\textit{V$_T$} with respect to the metal $E_F$ and transport in the device is thus by hot electrons. No additional bias is applied at the interface between the film and the substrate. After injection and transport through the LSMO thin film, the transmitted electrons are collected in the conduction band of Nb:STO using a third electrical contact. Only those electrons that retain sufficient energy and proper momentum to cross the LSMO/Nb:STO Schottky interface are collected\cite{Rippard, Tamalika Graphite,Tamalika PRL}. Hence the BEEM current, \textit{I$_B$}, is sensitive to scattering processes and gives important insights into the role of electron correlations to \textit{I$_B$} during transport.\\
\indent A typical \textit{I$_B$}-\textit{V$_T$} curve for hot electron transmission in LSMO at room temperature (RT) is shown in Fig. 2.  BEEM transmissions were recorded for four thicknesses of LSMO from 7 to 11 u.c. at different \textit{I$_T$} (Fig. 2a). Each BEEM spectra represents an average of at least 50 individual \textit{I$_B$} spectra, measured by positioning the STM tip at several different regions of the film. Approximately four devices of each thickness were measured. \textit{I$_B$} is observed to decrease with increasing thickness of LSMO. The transfer ratio of the collected current to the injected current (\textit{I$_B$}/\textit{I$_T$}) at -2 V for the 7 u.c. LSMO is 0.17 $\times$ 10$^{-3}$, and reduces to 0.01 $\times$ 10$^{-3}$ for the 11 u.c. LSMO. For all cases the sign of the current corresponds to electrons flowing from LSMO to Nb:STO and into the ohmic contact. An onset of \textit{I$_B$} is observed at around 1.06 $\pm$ 0.02 eV, which thereafter increases with increasing sample-tip bias . This corresponds to the local Schottky barrier height ($\phi_B$) extracted using the Bell-Kaiser model\cite{{beem}}, by plotting the square root of \textit{I$_B$} with sample-tip bias, \textit{V$_T$},  $[\frac{I_B}{I_T}\propto(V_T-\phi_B)^{2}]$ as shown in the inset of Fig. 2b.  A homogeneous distribution of the local $\phi_B$ in all devices is obtained which compares well with that obtained from macroscopic I-V, C-V and IPE measurements (see the Supplementary Information). We also observed an almost linear trend in \textit{I$_B$} with the injected tunnel current, \textit{I$_T$}, as shown in Fig. 2b, in accord with BEEM theory invoking planar tunneling\cite{LudekeBauer}.\\
\indent The collected BEEM current depends not only on the tunneling current injected into LSMO but also on the energy and momentum distribution of the carriers reaching the interface and the transmission probability at the LSMO/Nb:STO interface. Inelastic scattering, such as due to electron-electron interactions, can reduce the energy of the injected electron by $\sim$ 50 \%, whereas elastic scattering from impurities and defects or quasielastic scattering from acoustic phonons render the electron distribution isotropic. The epitaxial LSMO/Nb:STO interfaces studied here have been optimized to be fully strained, atomically abrupt, and with high crystalline perfection\cite{FittingK}. Thus they have fewer elastic scattering sites than the typical polycrystalline metal Schottky interfaces studied by BEEM, implying the conservation of transverse momentum of the transmitted electrons across the interface and with minimal influence on \textit{I$_B$}. \textit{I$_B$} also depends on the acceptance angle for electron collection at the Nb:STO interface which is determined by the ratio of the effective masses of the Nb:STO and LSMO. This is calculated to be larger here than found for most standard M-S interfaces (such as Au on Si).  Availability of allowed states in the conduction band minima (CBM) in $k$ space in Nb:STO is another criterion that further governs collection. Electronic band structure calculations\cite{STO Band structure} show that the projected conduction band minima in doped STO are at the zone center ($\Gamma$), thus electrons with small parallel momenta should be easily collected in the available phase space. Despite these favourable conditions, a central observation of this first application of BEEM to epitaxial perovskite heterostructures is the strong attenuation observed here, as compared to the highly disordered M-S structures previously studied by this technique. By way of comparison, \textit{I$_B$}/\textit{I$_T$} is almost 2 orders of magnitude higher for a similarly thick polycrystalline Ni film on a n-Si/Au M-S interface \cite{SubirPRB2012}. Thus we conclude that intrinsic correlation effects are dominant in the measurement.\\
\indent From the data in Fig. 2, we can extract the hot electron attenuation length, $\lambda$, in LSMO and study its energy dependence. For electron transmission in LSMO, $\lambda$ is obtained from $I_B(t,E)=I_B(0,E)exp[-t/{\lambda}(E)]$, where $E$ is the sample-tip bias and \textit{t} is the thickness of LSMO. From Fig. 3a, we find $\lambda$ in LSMO to be 1.48 $\pm$ 0.10 u.c. at -1.9 V and increases to 2.02 $\pm$ 0.16 u.c. at -1.3 V. Using Matthiessen's rule, the hot electron attenuation length, $\lambda (E)$, can be written as the sum of elastic $\lambda_{elastic}$ and inelastic $\lambda_{inelastic}(E)$ scattering lengths as:  
\begin{equation}
\frac{1}{\lambda(E)} = \frac{1}{\lambda_{elastic}} + \frac{1}{\lambda_{inelastic}(E)} 
\end{equation}
As argued above, elastic scattering is minimal at such epitaxial heterostructures and $\lambda (E)$ is thus dominated by $\lambda_{inelastic}(E)$. The energy dependence of $\lambda$ in LSMO is shown in Fig. 3b.\\
{\bf Discussions}\\
\indent Optimally doped LSMO is a ferromagnet and a transport half-metal with an insulating gap for minority spins at the Fermi level, $E_F$, and conducting for majority spins\cite {ParkNature}. However, at energies higher than $E_F$ and relevant for our studies, the density of states for both the majority and minority spin electrons increases\cite{PickettSingh1,PickettSingh2}. Conduction and ferromagnetism in LSMO is governed by the interaction of localized electrons from the incomplete $3d$ shell in Mn, by a process commonly referred to as the Zener double-exchange mechanism\cite{Zener}. Furthermore, as suggested by neutron scattering experiments\cite{Chen}, LSMO at room temperature consists of dynamic nanoscale polarons, which do not freeze out below the Curie temperature (as opposed to the situation in La$_{0.7}$Ca$_{0.3}$MnO$_{3}$). LSMO separates into regions where the electrons are trapped in a local Jahn-Teller distortion, and a conductive network without distortions\cite{Louca}. The transport then occurs as the hot electrons move through the conductive network, while being scattered by the local Jahn-Teller distortions, in addition to the electron-electron scattering of the injected charges themselves. Here, the polaron scattering can be regarded as static because the time scale of lattice motion (picoseconds) is much larger than that of hot electron transport through the LSMO film (femtoseconds)\cite{Matsuzaki}.\\
\indent We show that both mechanisms are present in our experiments in Fig. 3b. First, we have attempted a Fermi-liquid theory fit based on the density of states (DOS) previously reported. However, the experimentally determined scattering rate decreases slower than that predicted by this model. Incorporating an energy-independent fitting parameter that accounts for the polaronic scattering processes, we can fit our experimental data by the solid red curve ($\lambda_{e-e+polaron}$) in Fig. 3b. The obtained fitting constants are well within the ranges expected based on estimates of the band masses and DOS from density functional theory \cite{PickettSingh2}. The mean free path due to just the polaronic scattering is expected to be on the order of the polaron separation (or the polaron size). A previous study with neutron measurements estimated this value to be $\sim$2 u.c. (0.77 nm)\cite{Chen} which is roughly consistent with the constant value obtained here. Thus, the hot electron energy dependence enables us to quantitatively isolate two different scattering mechanisms: electron-electron (blue dotted curve) and polaronic scattering (green dotted line).\\
\indent Further, using the imaging capabilities in BEEM we visualize local hot-electron transport across a Nb:STO/LSMO heterointerface. STM topography of the LSMO (9 u.c.)/Nb:STO interface alongwith simultaneously recorded spatial map of the transmitted current, at \textit{V$_T$} = -2.5 V and \textit{I$_T$} = 8 nA are shown in Fig. 4. An atomically flat singly terminated TiO$_2$ surface with a step-height of 0.4 nm is observed from the STM topography whereas, a cross-section profile in the same location reveals different transmission at the terrace and at the step-edge with \textit{I$_T$} being constant. A histogram of the transmitted current at the terrace (area under the blue box in b) is shown in the inset in d. The mean value of \textit{I$_B$} matches well with that of the BEEM spectra for this film (Fig. 2). The reduction in \textit{I$_B$} at the step-edge as compared to that at the terrace arises due to the sensitivity of the propagating hot electrons to momentum scattering at the step-edges. This broadens the hot electron distribution and concomitantly reduces \textit{I$_B$} at such locations. This observation highlights the unique capability of the BEEM to study and directly visualize local electron transport in oxide heterointerfaces at the nanoscale.\\
\indent Our experimental method, based on hot electron transport in a vertical device structure of LSMO on Nb:STO, provides a first experimental measure of the hot electron attenuation length in a strongly correlated transition metal oxide as LSMO. This approach to probe electron transport, on the nanoscale, will open up exciting possibilities to both understand and tailor the electronic properties at oxide heterointerfaces and advance this emerging field of oxide electronics. \\

{\bf Methods}\\
\indent The devices were fabricated by pulsed laser deposition using TiO$_2$-terminated Nb:SrTiO$_3$ (001) substrates (Nb = 0.01 wt. $\%$). An SrMnO$_3$ single unit cell was first grown to enhance the SBH and suppress reverse bias leakage \cite{HikitaPRBSBH}, and subsequently LSMO films were grown at the O$_2$ partial pressure of 10$^{-1}$ Torr, the substrate temperature of 850$^{\circ}$ C, and the laser fluence of 0.8 J/cm$^2$. The deposited layer thicknesses were controlled by using reflection high-energy electron diffraction intensity oscillations. For Ohmic contacts, gold was evaporated onto the LSMO, and indium was ultrasonically soldered onto the Nb:STO. A modified commercial STM system from RHK Technology was used for the BEEM studies. All the BEEM measurements were performed at 300 K using PtIr metal tips, in the constant current mode. 
A Pt wire was used to ground the top metal contact. Indium solder defined the Ohmic contact.
Typically 10 devices were fabricated for each thickness. Each BEEM spectra represents 
an average of at least 50 individual \textit{I$_B$} spectra, measured by positioning the STM 
tip at several different regions of the film. Approximately four devices of each thickness were measured. The BEEM current is detected with a two-stage amplifier (10$^{11}$ V/A). \\

\clearpage

\begin{figure}[htb]
\includegraphics[scale=0.55]{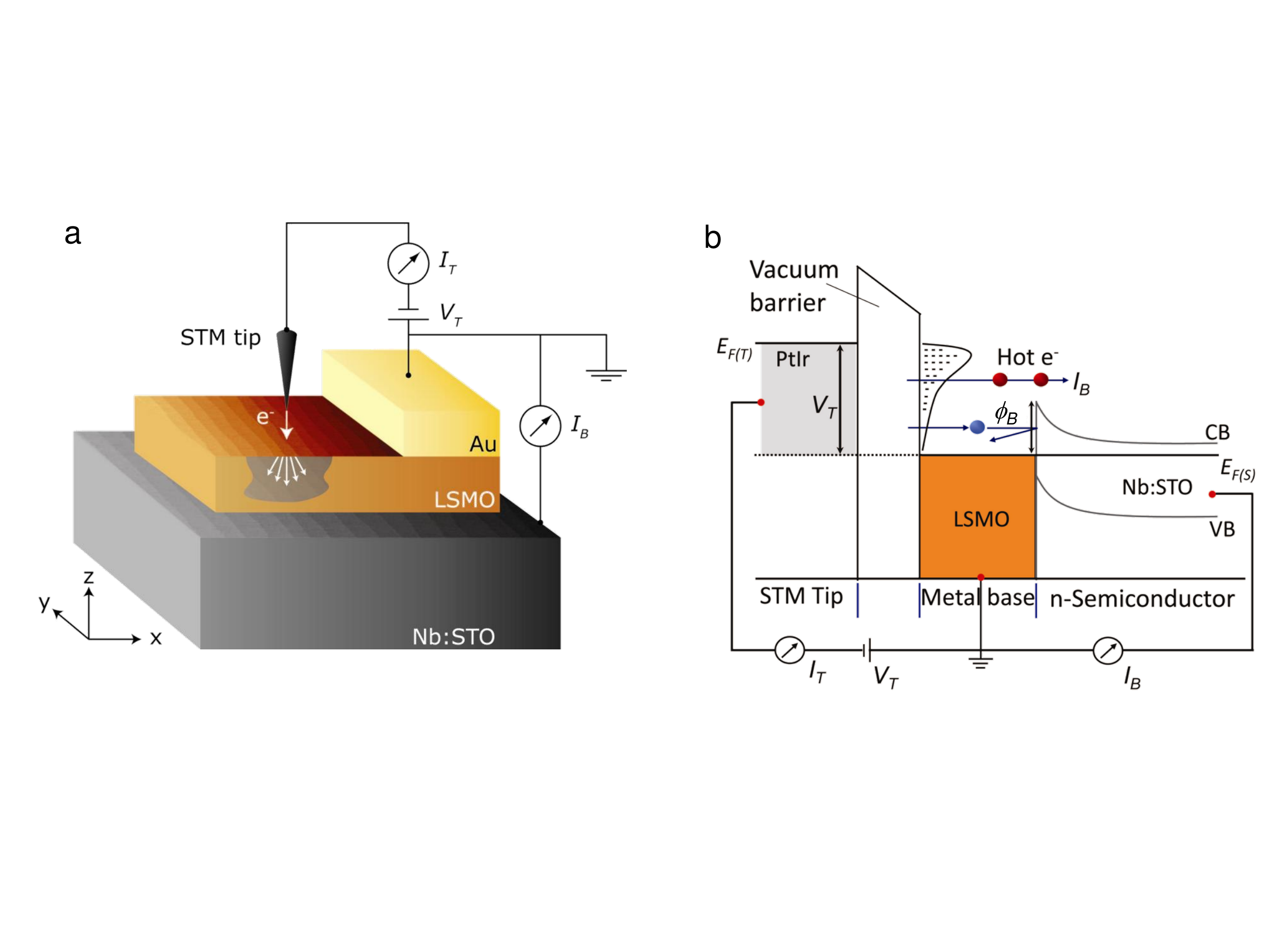}
\caption{\label{Figure1}{\bf Principle of the BEEM experiment.} {\bf a,} Schematics of the BEEM technique. The
sample is a thin epitaxial film of LSMO of variable thickness deposited on an isostructural 0.01 wt. $\%$ Nb:STO substrate. A PtIr STM tip is used to locally inject electrons into the sample by tunneling at a sample-tip bias, \textit{V$_T$}, between the tip and the LSMO surface. The electrons transmitted perpendicularly through the LSMO layer are collected as \textit{I$_B$} in the Nb:STO with a third (rear) electrical contact. {\bf b,} Energy schematics of the BEEM technique. Hot electrons are emitted from the STM tip across the vacuum tunnel barrier and injected locally into the LSMO base.  After transmission in the base, they are then collected in the conduction band of the Nb:STO, provided the energy and momentum criteria needed to overcome $\phi_B$ at the Nb:STO interface are satisfied (red circles; blue circle denotes those electrons which do not satisfy the criteria). All measurements were performed at 300 K.} 
\end{figure}

\clearpage
\begin{figure}[htb]
\includegraphics[scale=0.55]{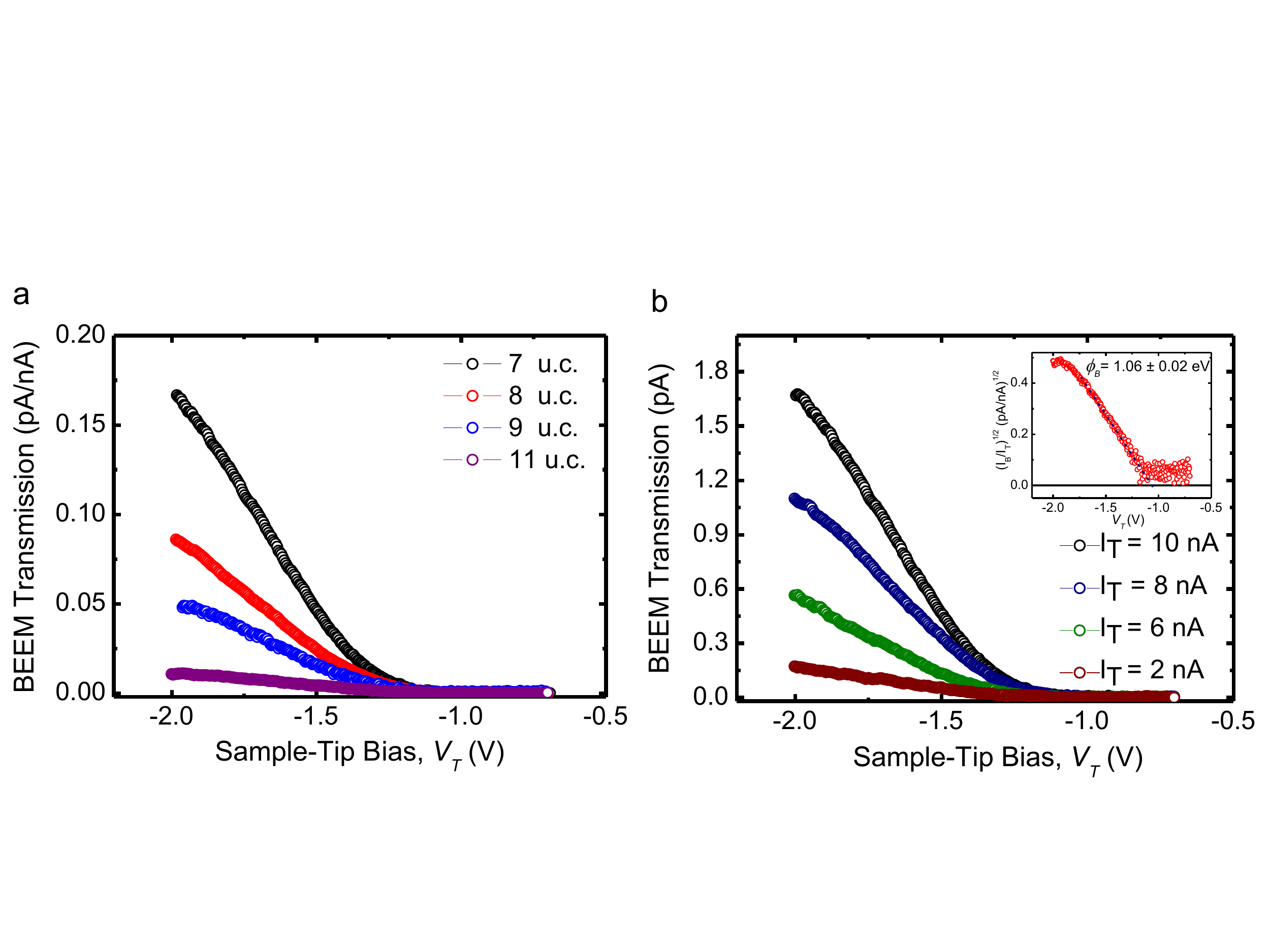}
\caption{\label{Figure2}{\bf Energy dependence of hot electron transmission in LSMO at the nanoscale.} {\bf a,} Electron current per nA of injected tunnel current versus sample-tip bias across LSMO/Nb:STO (001), for LSMO thickness, \textit{t}, of  7 u.c. (black), 8 u.c. (red), 9 u.c. (blue) and 11 u.c. (purple). The sign of the current corresponds to electrons flowing from the LSMO layer to the Nb:STO and into the ohmic contact. Typically 10 devices were fabricated for each thickness. Each curve represents an average of over 50 different spectra collected at different regions, in approximately 4 devices of each thickness. BEEM transmission decreases with increasing thickness of LSMO.  With an increase in sample-tip bias an onset of \textit{I$_B$} at around 1.06 $\pm$ 0.02 eV, corresponding to $\phi_B$ is seen. {\bf b,} \textit{I$_B$} is observed to increase with tunnel current \textit{I$_T$}, as is shown for a LSMO (7 u.c.)/Nb:STO (001) device. The collected current is shown for tunnel currents of 2 nA (brown), 6 nA (green), 8 nA (blue) and 10 nA (black). Inset shows the extracted $\phi_B$ for LSMO (7 u.c.)/Nb:STO, obtained by fitting to the Bell-Kaiser model, typically used for fitting BEEM spectra. The local $\phi_B$ is extracted by plotting the square root of the normalized \textit{I$_B$} with \textit{V$_T$}.}  
\end{figure}

\clearpage

\begin{figure}[t]
\includegraphics[scale=0.55]{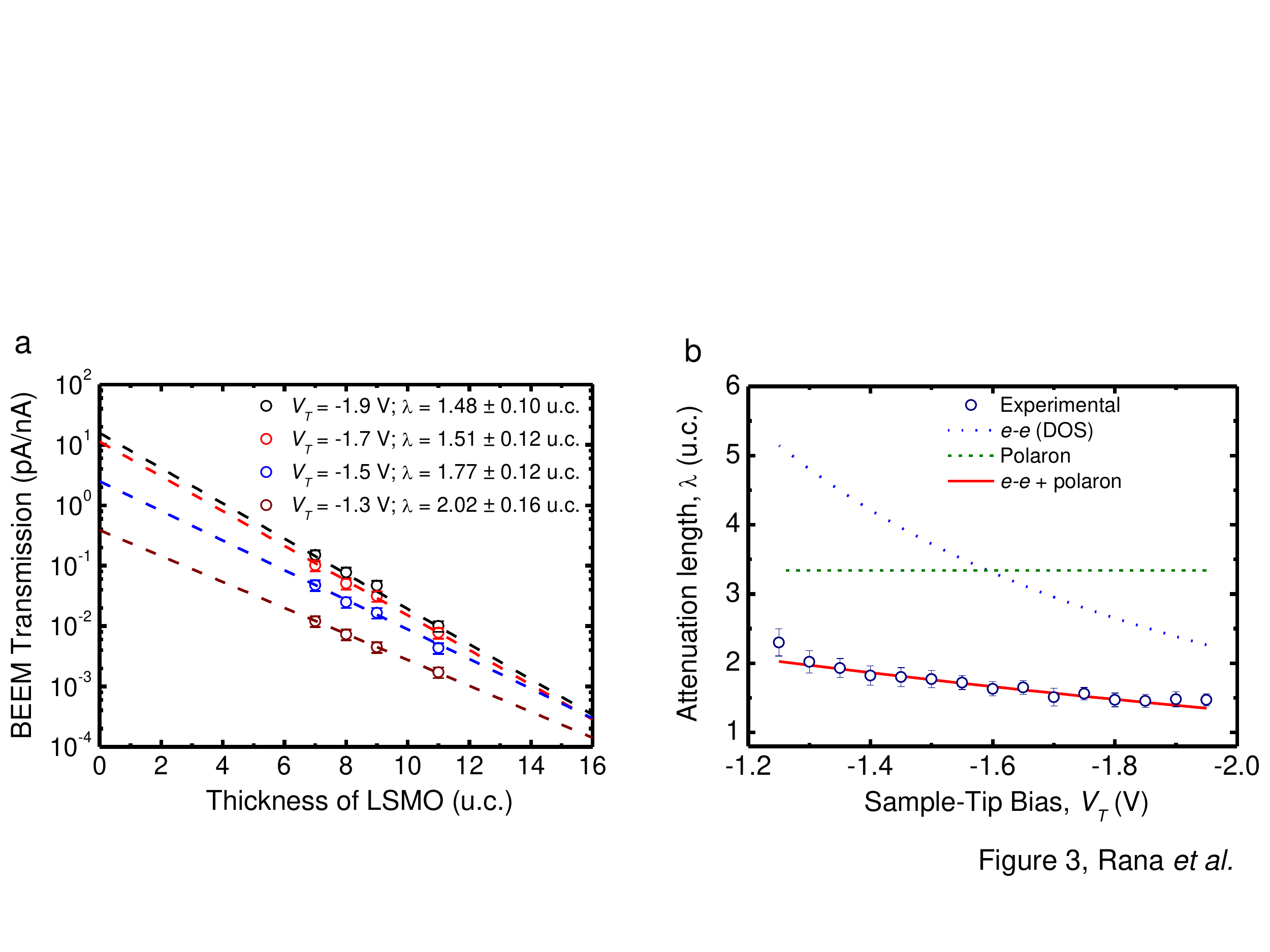}
\caption{\label{Figure3}{\bf Energy dependence of the hot electron attenuation length in LSMO.} {\bf a,}
BEEM transmission normalized per nA of injected tunnel current versus the thickness of LSMO obtained from the data in Fig. 2a. This is plotted for sample-tip bias of -1.9 V (black circles), -1.7 V (red circles), -1.5 V (blue circles) and -1.3 V (brown circles). Dotted lines at each energy represent exponential decay of \textit{I$_B$} with attenuation length of 1.48 $\pm$ 0.10 u.c., 1.51 $\pm$ 0.12 u.c., 1.77 $\pm$ 0.12 u.c. and 2.02 $\pm$ 0.16 u.c. at these energies. Error bars represent the deviation from the best exponential fit. These lines can be extrapolated to zero LSMO thickness, which corresponds to attenuation due to the interface. {\bf b,} Attenuation length, $\lambda$, extracted from Fig. 2a for different energy values. The solid red curve which fits the experimental data is a theoretical
estimate taking into account electron-electron scattering within Fermi liquid theory based on the DOS for the materials and an
energy-independent (at the energies shown) polaron scattering term. The dotted curves depict the two contributions independently, where
blue (green) shows the electron-electron (polaron) contribution to the full estimate.}  
\end{figure}

\begin{figure}[t]
\includegraphics[scale=0.55]{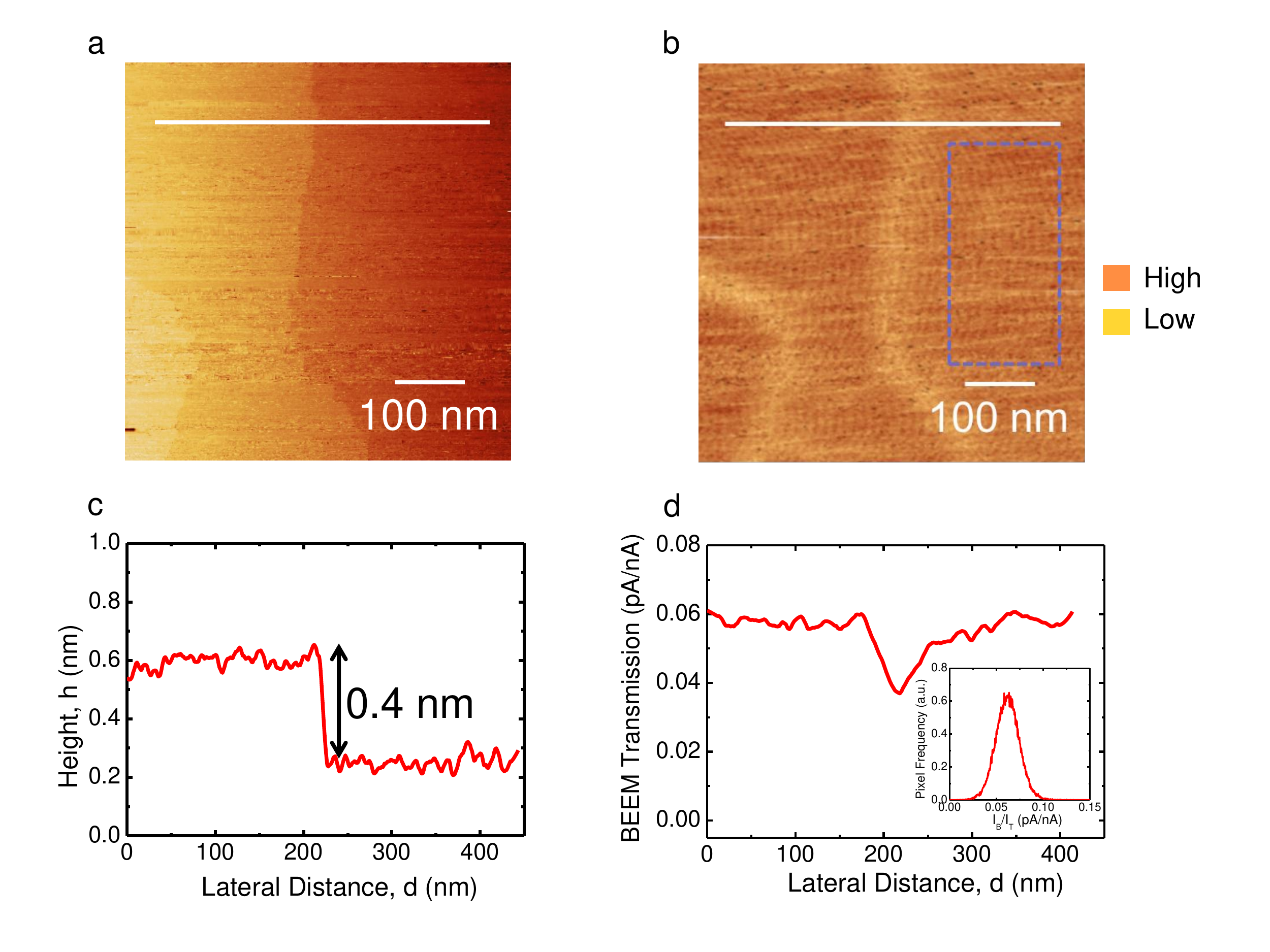}
\caption{\label{Figure4}{\bf Spatially resolved electron transmission through
LSMO (9 u.c.)/Nb:STO epitaxial heterostructure.} {\bf a, c}, Topographic STM image taken at \textit{V$_T$}= -2.5 V and \textit{I$_T$}= 8 nA over a 0.45$\times$0.45 $\mu$m$^{2}$ area. Epitaxial LSMO layer grown on atomically flat singly terminated  TiO$_2$ surface is confirmed from the 0.4 nm step-height, as seen in the cross-section profile in c along the white line in a. {\bf b, d}, corresponding map of the BEEM current, \textit{I$_B$}, acquired simultaneously. The BEEM transmission is seen to be reduced by a factor of 2 while approaching a step-edge from a terrace as indicated by a dip across the cross-section profile in d. Quantitative analysis of \textit{I$_B$} using a histogram of current distribution in the area under the blue box in b is shown in the inset of d. The mean value of the distribution matches well with that of the BEEM transmission of this device as shown in Fig.2 (left).}  
\end{figure}
\clearpage

\clearpage
\renewcommand\refname{R\lowercase{eferences}}

\clearpage

\noindent {\bf Acknowledgements}\\
We acknowledge financial support from the NWO-VIDI program and the NanoNed program coordinated by the Dutch Ministry of Economic Affairs (K.G.R, S.P \& T.B) and thank R. Ruiter for help with an illustration. We acknowledge support from the Department of Energy, Office of Basic Energy Sciences, Division of Materials Sciences and Engineering, under contract DE-AC02-76SF00515 (T.Y., A.F.K., T.P.D., Y.H., \& H.Y.H.). T.Y. also acknowledges support from the Japan Society for the Promotion of Science (JSPS). \\

\noindent {\bf Authors contributions}\\
T.Y. performed the device fabrication, the current-voltage, capacitance-voltage measurements, and the internal photoemission spectroscopy. K.G.R. performed all the BEEM measurements along with S.P. and contributed to the analysis of the data ; theoretical fits to the extracted data were performed by A.F.K. and T.P.D. Y.H. and H.Y.H. assisted with the planning and the measurements. T.B. conceived and supervised the research and contributed to the analysis of the data. All co authors extensively discussed the results and provided important insights. T.B. wrote the manuscript with input from all co authors.\\

\noindent {\bf Additional Information}\\
Supplementary Information accompanies this paper at http://www.nature.com/scientificreports
The authors declare no competing financial interest.\\

\clearpage


\begin{thebibliography}{99}
\bibitem{HwangTokura} Hwang, H. Y., Iwasa, Y., Kawasaki, M., Keimer, B., Nagaosa, N., \& Tokura, Y. Emergent phenomena at oxide interfaces. {Nature Mater.} {\bf 11,} 103-113 (2012).
\bibitem{CheongPV} Choi, T., Lee, S., Choi, Y. J., Kiryukhin, V. \& Cheong, S.-W. Switchable ferroelectric diode and photovoltaic effect in BiFeO$_3$. {Science} {\bf 324,} 63-66 (2009).
\bibitem{RameshNatNanotech2010}Yang, S. Y. {et al.} Above-bandgap voltages from ferroelectric photovoltaic devices. {Nature Nanotech.} {\bf 5,} 143-147 (2010).
\bibitem{Takeaki} Yajima, T., Hikita, Y. \& Hwang, H. Y. A heteroepitaxial perovskite metal-base transistor. {Nature Mater.} {\bf10,} 198-201 (2011).
\bibitem{Garcia2009} Garcia, V., Fusil, S., Bouzehouane, K., Enouz-Vedrenne, S., Mathur, N. D., Barth\'{e}l\'{e}my, A., \& Bibes, M. Giant tunnel electroresistance for non-destructive readout of ferroelectric states. {Nature} {\bf 460,} 81-84 (2009).
\bibitem{Maksymovych} Maksymovych, P., Jesse, S., Ramesh, R., Baddorf, A. P., \& Kalinin, S. V. Polarization control of electron tunneling into ferroelectric surfaces. {Science} {\bf 324,} 1421-1425 (2009).
\bibitem{Aeschlimann} Aeschlimann, M., Bauer, M., Pawlik, S., Weber, W., Burgermeister, R., Oberli, D., \& Siegmann, H. Ultrafast spin-dependent electron dynamics in fcc Co. {Phys. Rev. Lett.} {\bf 79,} 5158-5161 (1997). 
\bibitem{MonsmaScience} Monsma, D. J., Vlutters, R., \& Lodder, J. C.  Room temperature-operating spin-valve transistors formed by vacuum bonding. {Science} {\bf 281,} 407-409 (1998).
\bibitem{BuhrmanSankey} Sankey, J. C., Cui, Y-T., Sun, J. Z., Slonczewski, J. C., Buhrman,  R. A., \& Ralph, D. C. Measurement of the spin-transfer-torque vector in magnetic tunnel junctions. {Nature Phys.} {\bf 4,} 67-71 (2008).
\bibitem{AppelbaumNature} Appelbaum, I., Huang, B., \& Monsma, D. J. Electronic measurement and control of spin transport in silicon. {Nature} {\bf 447,} 295-298 (2007).
\bibitem{Gabor} Gabor, N. M., Song, J. C. W., Ma, Q., Nair, N. L., Taychatanapat, T., Watanabe, K., Taniguchi, T., Levitov, L. S., \& Jarillo-Herrero, P. Hot carrier-assisted intrinsic photoresponse in Graphene. {Science} {\bf 334,} 648-652 (2011).
\bibitem{beem} Kaiser, W. J., \& Bell, L. D. Direct investigation of subsurface interface electronic structure by Ballistic-Electron-Emission Microscopy.  {Phys. Rev. Lett.} {\bf 60,} 1406-1409 (1988). 
\bibitem{HikitaBook} Hikita, Y. and Hwang, H. Y. ``Complex Oxide Schottky Junctions" in Thin Film Metal-Oxide Fundamental and Applications in Electronics and Energy, ed. S. Ramanathan, Springer 2010.
\bibitem{vonkanel1} Meyer, T., \& von K\"{a}nel, H. Study of interfacial point defects by Ballistic Electron Emission Microscopy. {Phys. Rev. Lett.} {\bf 78,} 3133-3136 (1997).
\bibitem{HikitaPRBSBH} Hikita, Y., Nishikawa, M., Yajima, T., \& Hwang, H. Y. Termination control of the interface dipole in La$_{0.7}$Sr$_{0.3}$MnO$_{3}$/Nb:SrTiO$_3$ (001) Schottky junctions.  {Phys. Rev. B} {\bf79,} 073101 (2009).
\bibitem{Rippard} Rippard, W. H., \& Buhrman,  R. A. Ballistic electron magnetic microscopy: Imaging magnetic domains with nanometer resolution. {Appl. Phys. Lett.} {\bf 75,} 1001-1003 (1999).
\bibitem{Tamalika Graphite} Banerjee, T., van der Wiel, W. G., \& Jansen, R. Spin injection and perpendicular spin transport in graphite nanostructures. {Phys. Rev. B} {\bf 81,} 214409 (2010).
\bibitem{Tamalika PRL} Banerjee, T., Haq, E., Siekman, M. H., Lodder, J. C., \& Jansen, R. Spin filtering of hot holes in a metallic Ferromagnet. {Phys. Rev. Lett.} {\bf 94,} 027204 (2005).
\bibitem{LudekeBauer} Ludeke, R., \& Bauer, A. Hot electron scattering processes in metal films and at metal-semiconductor interfaces. {Phys. Rev. Lett.} {\bf 71,} 1760-1763 (1993).
\bibitem{FittingK} Fitting Kourkoutis, L., Song, J. H., Hwang, H. Y., \& Muller, D. A. Microscopic origins for stabilizing room-temperature ferromagnetism in ultrathin manganite layers.  {Proc. Natl. Acad. Sci. U.S.A.} {\bf 107,} 11682-11685 (2010).
\bibitem{STO Band structure} Mattheiss, L. F. Energy Bands for KNiF$_3$, SrTiO$_3$, KMoO$_3$, and KTaO$_3$. {Phys. Rev. B} {\bf 6,} 4718-4740 (1972).
\bibitem{SubirPRB2012} Parui, S., Rana, K. G., Bignardi, L., Rudolf, P., van Wees, B. J., \& Banerjee, T. Comparison of hot-electron transmission in ferromagnetic Ni on epitaxial and polycrystalline Schottky interfaces.  {Phys. Rev. B} {\bf 85,} 235416 (2012).
\bibitem{ParkNature} Park, J. -H., Vescovo, E., Kim, H. -J., Kwon, C., Ramesh, R., \& Venkatesan, T. Direct evidence for a half-metallic ferromagnet. {Nature} {\bf 392,} 794-796 (1998).
\bibitem{PickettSingh1} Pickett, W. E., \& Singh, D. J. Chemical disorder and charge transport in ferromagnetic manganites.{Phys. Rev. B} {\bf 55,} R8642-R8645 (1997).
\bibitem{PickettSingh2}Pickett, W. E., \& Singh, D. J. Electronic structure and half-metallic transport in the La$_{1-x}$Ca$_x$MnO$_3$ system. {Phys. Rev. B} {\bf 53,} 1146-1160 (1996).
\bibitem{Zener} Zener, C. Interaction between the d-Shells in the transistion metals. II ferromagnetic compounds of manganese with perovskite structure. {Phys. Rev.} {\bf 82,} 403-405 (1951).
\bibitem{Chen} Chen, Y., Ueland, B. G., Lynn, J. W., Bychkov, G. L., \& Barilo, S. N. Polaron formation in the optimally doped ferromagnetic manganites La$_{0.7}$Ca$_{0.3}$MnO$_{3}$ and La$_{0.7}$Ba$_{0.3}$MnO$_{3}$. {Phys. Rev. B} {\bf 78,} 212301 (2008).
\bibitem{Louca} Louca, D., \& Egami, T. Local lattice distortions in La$_{1 - x}$Sr$_{x}$MnO$_{3}$ studied by pulsed neutron scattering. {Phys. Rev. B} {\bf 59,} 6193-6204 (1999).
\bibitem{Matsuzaki} Matsuzaki, H., Uemura, H., Matsubara, M., Kimura, T., Tokura, Y., \& Okamoto, H. Detecting charge and lattice dynamics in photoinduced charge-order melting in perovskite-type manganites using a 30-femtosecond time resolution. {Phys. Rev. B} {\bf 79,} 235131 (2009).

 
\end{thebibliography}
\end{document}